\documentclass[preprint]{aastex63}       

\usepackage{longtable} 
\usepackage{threeparttable} 
\usepackage{booktabs}    
\usepackage{bibunits}
\usepackage{bm}
\usepackage{amsmath, amssymb}
\usepackage{lineno}
\usepackage{rotating}

\accepted{for publication in ApJ}

\shortauthors{Hou et al.}

\begin{document}
\title{Deep search for gamma-ray emission from the accreting X-ray pulsar 1A~0535+262} 

\correspondingauthor{X. Hou, D.F. Torres}
\email{xhou@ynao.ac.cn, dtorres@ice.csic.es}

\author[0000-0003-0933-6101]{X. Hou}
\affiliation{Yunnan Observatories, Chinese Academy of Sciences, Kunming 650216, China}
\affiliation{Key Laboratory for the Structure and Evolution of Celestial Objects, Chinese Academy of Sciences, Kunming 650216, China}

\author[0000-0003-2839-1325]{W. Zhang}
\affiliation{Institute of Space Sciences (ICE, CSIC), Campus UAB, 08193 Barcelona, Spain}
\affiliation{Institut d’Estudis Espacials de Catalunya (IEEC), 08034 Barcelona, Spain}

\author[0000-0002-1522-9065]{D.F. Torres}
\affiliation{Institute of Space Sciences (ICE, CSIC), Campus UAB, 08193 Barcelona, Spain}
\affiliation{Institut d’Estudis Espacials de Catalunya (IEEC), 08034 Barcelona, Spain}
\affiliation{Institució Catalana de Recerca i Estudis Avançats (ICREA), E-08010 Barcelona, Spain}

\author[0000-0001-9599-7285]{L. Ji}
\affiliation{School of Physics and Astronomy, Sun Yat-Sen University, Zhuhai 519082, China}

\author[0000-0003-1720-9727]{J. Li}
\affiliation{CAS Key Laboratory for Research in Galaxies and Cosmology, Department of Astronomy, University of Science and Technology of China, Hefei, China}
\affiliation{School of Astronomy and Space Science, University of Science and Technology of China, Hefei, China}

\begin{abstract}

Binary systems are a well-established subclass of gamma-ray sources. The high mass X-ray binary pulsar 1A~0535+262 has been considered to be a possible gamma-ray emitter for a long time, although former gamma-ray searches using \textit{Fermi}-LAT and VERITAS data resulted in upper limits only. We aim at a deep search for gamma-ray emission and pulsations from 1A~0535+262 using more than 13 years of \textit{Fermi}-LAT data. The analysis was performed for both the whole \textit{Fermi}-LAT data set, as well as for the X-ray outbursts that 1A~0535+262 has experienced since the launch of \textit{Fermi}. Various X-ray observations have been used to generate the ephemeris for the pulsation search. We also investigate the long-term gamma-ray flux variability and perform orbital phase-resolved analysis for the outbursts. We did not detect any steady or pulsed gamma-ray emission from 1A~0535+262 during the whole \textit{Fermi}-LAT mission span or its X-ray outbursts. We thus derived the deepest gamma-ray luminosity upper limits to date at the 95\% confidence level to be around (2.3$-$4.7)$\times 10^{32}\, \rm erg \, s^{-1}$ depending on different spectral indices assumed, which results in a ratio of $L_{\rm \gamma}$ to $L_{\rm X}$ (2$-$150 keV) being (1.9$-$3.9)$\times10^{-6}$.
\end{abstract}


\section{Introduction}
Among the various gamma-ray sources detected in the MeV/GeV and/or even the TeV band, binary systems are a well-established subclass, although their number is yet small.
Interestingly, despite their small number, several varieties of binaries exist with different gamma-ray emission mechanisms. 
For a recent review, see \cite{Dubus2015}. 
First, \textit{Fermi}-LAT has detected gamma-ray binaries themselves, usually defined as a subclass of high mass X-ray binaries (HMXBs) with O or B companion star, with two main features:
they emit modulated gamma rays peaking above 1 MeV and present orbital variability at all frequencies.
The spectral energy distribution is usually thought to be powered by the pulsar/stellar wind interaction, although so far the central compact object has been associated to known pulsars only in three cases: PSR B1259-63/LS 2883
\citep[see][and references therein]{HESS2020,Chernyakova2020a}, PSR J2032+4127/MT91 213 \citep[see][and references therein]{Coe2019},
and the recent identification of LS I +61 303 \citep{Weng2022}. 

Recycled pulsars in binaries, called redbacks \& black widow systems, are tight (orbital periods less than a day), low mass ($< 0.1 M_\odot$) gamma-ray binaries with a main sequence degenerate companion. 
In these systems the  pulsar wind is ablating the companion star, leading to eclipses, radio variability, X-ray/radio anti-correlation, and pulsar nulling \citep[see, e.g.,][]{Bogdanov2018}. 
Transitional pulsars are special among these systems, since they exhibit two different states (accretion and rotation powered) which may interchange in a matter of weeks and persist for years \citep[see e.g.,][]{Archibald2009,Papitto2013,deMartino2015}. 
These state changes produce significant gamma-ray variability \citep{Stappers2014,Torres2017}. 

Other \textit{Fermi}-LAT detected binaries include microquasars, for which the gamma rays seem to be associated with a relativistic jet. Notable examples are Cyg X-1 \citep[see e.g.,][]{Albert2007,Zanin2016,Zdziarski2017} and Cyg X-3 \citep[see e.g.,][]{Abdo2009c,Tavani2009,Zdziarski2018,Sinitsyna2019}. 
%
%
The microquasar SS 433, whose central object is still undetected, was unexpectedly found to produce gamma rays far from the jet \citep{HAWC2018,Rasul2019,Xing2019a,Fang2020}
 with a GeV source showing variability with the precession period of the system \citep{Li2020}.

%
Finally, one finds the accreting millisecond pulsar SAX J1808.4$-$3658 \citep{Emma2016}: a ``\textit{bona fide}'' accreting system apparently emitting gamma rays, although yet not significantly.
This is the only such system for which a gamma-ray detection was hinted so far.
Putting upper limits on accreting sources or detecting them is a must to understand the different variety.

1A~0535+262 is one of the best studied HMXB accreting pulsars. It was discovered in 1975 by the Rotation Modulation Collimator on \textit{Ariel V}, with a pulsation period of 104 s \citep{Rosenberg1975}. 
The compact object in the system is a highly magnetized neutron star which accretes mass from the O9.7IIIe companion star \citep{Steele1998}. 
The orbital period of the system is $\sim 111$ days \citep{Coe2006} and the eccentricity of the orbit is $e=0.47\pm0.02$ \citep{Finger1996}. 
1A~0535+262 is relatively close to Earth with a distance of $1.8\pm0.1$ kpc, as measured by \textit{Gaia} \citep{Bailer2018}. 
Since its discovery, 1A~0535+262 has exhibited different X-ray outbursts with peak flux ranging from $\sim100$ mCrab to $\sim12.5$ Crab. In particular, three giant X-ray outbursts were detected since the launch of the {\it Fermi} satellite: in 2009 December \citep{Acciari2011}, 2011 February \citep{Sartore2015} and 2020 November \citep[][and references therein]{Kong2022,Mandal2022}. Figure \ref{fig:lc} shows the long-term light curves of 1A 0535+262 in different energy bands obtained from the \textit{Swift}/BAT and \textit{MAXI}/GSC Broadband Transient Monitor\footnote{\url{http://sakamotoagu.mydns.jp/bat_gsc_trans_mon/web_lc/1_Day.php?name=1A_0535+262}}. There was also a double-peaked outburst \citep{Caballero2013} just prior to the 2009 giant one, which is, however, not included in the monitor database.
VLA observations during the 2020 outburst revealed non-thermal radio emission from the source position \citep{Eijnden2020}.

1A~0535+262 was earlier associated with the EGRET unidentified gamma-ray source 3EG J0542+2610 \citep{Romero2001} and thus has long been considered as a high-energy (HE; $E >100$ MeV) and very-high-energy (VHE; $E >100$ GeV) emitter candidate. 
The first gamma-ray search for 1A~0535+262 dates back to more than 10 years, during its giant 2009 outburst.
On that occasion, the X-ray outburst in December of 2009 triggered VHE VERITAS observations.
Only upper limits have been derived \citep{Acciari2011}. 
These authors also did a search for HE gamma-ray emission from 1A~0535+262 with {\it Fermi}-LAT
in a period  spanning the onset of the X-ray outburst to the successive apastron of the pulsar (2009 November 30 to 2010 February 22). 
No significant GeV excess was seen and a flux upper limit of $F(>0.2{\,\rm GeV})<1.9 \times 10^{-8}$ photons cm$^{-2}$ s$^{-1}$ at 99\% confidence level was imposed.
Recently, \cite{Lundy2021} updated the VERITAS VHE search for 1A~0535+262 during its 2020 giant outburst. Again, only upper limits have been obtained. 
Also, \cite{Harvey2022} used 12.5 years LAT data to search for gamma-ray emission from 1A~0535+262.
They claimed a marginal persistent gamma-ray excess (3.5$\sigma$) at the position of the source and found that the gamma-ray activity may be correlated with the X-ray outbursts. 
In addition, they found that essentially all of the gamma-ray excess is concentrated in the orbital phase bin preceding periastron, thus providing evidence of the gamma-ray excess originating from this binary system. If real, these hints are relevant, and thus can help to get insights to the particle acceleration and emission process during the accretion. 
%

In this work, we analyze the three giant outbursts, in 2009, 2011 and 2020, and the previous double-peaked outburst of 1A~0535+262. The time span of each outburst was defined by investigating the X-ray light curves presented in the literature \citep{Acciari2011,Sartore2015,Mandal2022,Kong2022}. 
We perform a deep search for gamma-ray emission and pulsations from 1A 0535+262 using more than 13 years of {\it Fermi}-LAT data and the latest 12-year source catalog. We use the latest Instrument Response Functions (IRFs) and background diffuse models. 
Our work therefore extends the result presented in \cite{Harvey2022}. The paper is organized as follows: we describe the data analysis procedure and results in Section \ref{result}, and discuss our findings in Section \ref{discuss}. We finally conclude in Section \ref{conclude}.

\section{Data analysis and results}
\label{result}

\subsection{Timing solutions}

For the 2009 double-peaked and giant outbursts, we adopted the spin measurements and the orbital ephemeris of 1A 0535+262 from the {\it Fermi}/GBM monitoring\footnote{\url{https://gammaray.nsstc.nasa.gov/gbm/science/pulsars/lightcurves/a0535.html}}. We used a polynomial function to describe the spin evolution approximately.
For the 2011 outburst, we used the timing solution reported in \cite{Sartore2015} derived using \textit{INTEGRAL} observations. 
For the recent 2020 outburst, thanks to the extensive coverage of {\it Insight}-HXMT observations \citep{Wang2022}, we used the phase-connection technique \citep{Deeter1981} to determine the spin evolution accurately.
In practice, for each 1000\,s segment we folded background-subtracted light curves in the energy range of 25$-$80\,keV, for which the pulse profile shape is relatively stable.
The 1000\,s was chosen because this is the typical interval for \textit{Insight}-HXMT's good time.
The time-of-arrival (TOA) of each segment was estimated by cross-correlating these pulse profiles with an averaged template.
Then the spin evolution was determined by using the software {\sc Tempo2} \citep{Hobbs2006}.
We summarize the timing solutions for different outbursts used in the following pulsation search (Section \ref{pulsation}) in Table~\ref{tab:spin}.


\subsection{\textit{Fermi}-LAT data set and reduction}

We used the Pass 8 data set \citep{pass8Atwood,Bruel2018} available at the \textit{Fermi} Science Support Center (FSSC)\footnote{http://fermi.gsfc.nasa.gov/ssc/}.
This spans 166 months, from 2008 August 4 to 2022 June 9, with reconstructed energy in the range 0.1$-$300 GeV. 
We selected SOURCE class events (Front and Back) with a zenith angle smaller than $90^\circ$ to avoid the Earth limb contamination. 
The events were further filtered based on the criteria \texttt{``DATA\_QUAL>0 \&\& LAT\_CONFIG==1''} to get the good time intervals 
in which the satellite was working in standard data taking mode and the data quality was good.
We did not apply a Region of Interest (ROI)-based zenith angle cut\footnote{https://fermi.gsfc.nasa.gov/ssc/data/analysis/scitools/data\_preparation.html}. 
The data set was centered at 1A~0535+262 with coordinates $(\alpha,\delta)=(84\fdg7274,26\fdg3158)$, with a radius of $10^\circ$. 
The coordinates of 1A~0535+262 were taken from the SIMBAD\footnote{http://simbad.u-strasbg.fr/simbad/} database and in the J2000 frame. 
The analysis was performed using the P8R3\_SOURCE\_V3 IRFs and the latest Fermitools\footnote{https://fermi.gsfc.nasa.gov/ssc/data/analysis/software/} (v2.2.0) available at the FSSC.

\subsection{\textit{Fermi}-LAT spectral analysis}

In the spectral analysis, the latest 4FGL-DR3\footnote{https://fermi.gsfc.nasa.gov/ssc/data/access/lat/12yr\_catalog/} (gll\_psc\_v30.fit) \citep{4DFL-DR3} sources within a $20^\circ$ circle around 1A~0535+262 were included to build a complete spatial and spectral source model. 
We also included the latest Galactic interstellar emission model, ``gll\_iem\_v07.fits'', as well as the isotropic emission spectrum ``iso\_P8R3\_SOURCE\_V3\_v1.txt'',
with the latter taking into account the extragalactic emission and the residual instrumental background\footnote{https://fermi.gsfc.nasa.gov/ssc/data/access/lat/BackgroundModels.html}. 
Both the normalizations and spectral indices of sources within $5^\circ$ around 1A~0535+262 were set free to vary except for 4FGL J0534.5+2201i, which is recommended to be fixed to account for the Inverse Compton Scattering component of the Crab Nebula. 
Extended sources were modeled using the 12-year templates.
Since the closest source in the model is $\sim1.6^\circ$ away from 1A~0535+262, we added 1A~0535+262 manually in the model as a point source with a simple Power Law spectral model.
This allows us to check whether the addition of such source is significant.

Model fitting was performed in a $14^\circ \times 14^\circ$ ROI using the maximum likelihood method \citep{mattox96}. 
We followed the binned likelihood procedure outlined in the FSSC using a $0\fdg1 \times 0\fdg1$ pixel size and thirty logarithmic energy bins over 0.1$-$300 GeV. Two extra energy bins have been added to take into account the energy dispersion except for the isotropic component.
The significance of a given source in the model is characterized by the Test Statistic (TS), which is expressed as TS $=2(\log \mathcal{L}-\log \mathcal{L}_{0})$, where $\log \mathcal{L}$ and $\log \mathcal{L}_{0}$ are the logarithms of the maximum likelihood of the complete source model and of the background model (i.e. the source model without the given source included), respectively.

We first performed a global binned likelihood fit to the whole data set by fixing the spectral index of 1A~0535+262 to 2, 2.3 and 3, respectively.
Such spectral indices are chosen to represent possible emission mechanisms. 
At the same flux level, if the source were to emit a hard spectrum (as the assumed 2) across the {\it Fermi}-LAT energy regime, it should be easier to detect it due to the diminishing background at higher energies.
Then, using the best-fit source model from the whole data set fit, we performed binned likelihood fits to the different X-ray outbursts that 1A~0535+262 has experienced in the past, following the same fitting setup and procedure as for the whole data set. 
To increase the detection possibility and statistics, we also stacked all the outbursts together to perform the fit. 
1A~0535+262 was not detected in any of these cases and we therefore computed a 95\% confidence level energy flux upper limit accordingly. 
The fitting results are presented in Table \ref{tab:fermiflux}.

\subsection{Gamma-ray variability}
\label{variability}

\begin{figure*}
\centering
\includegraphics[scale=0.45]{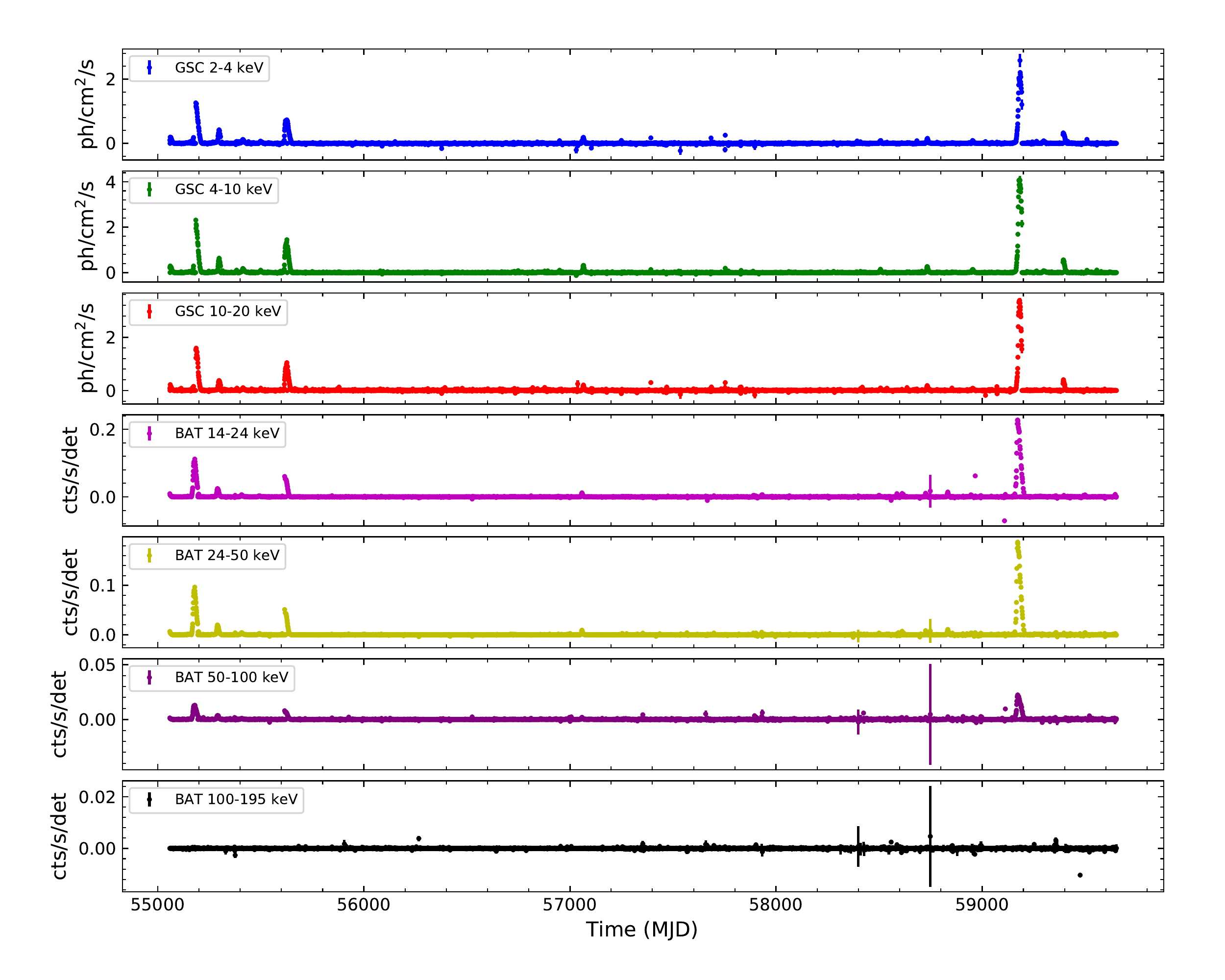}
\includegraphics[scale=0.5]{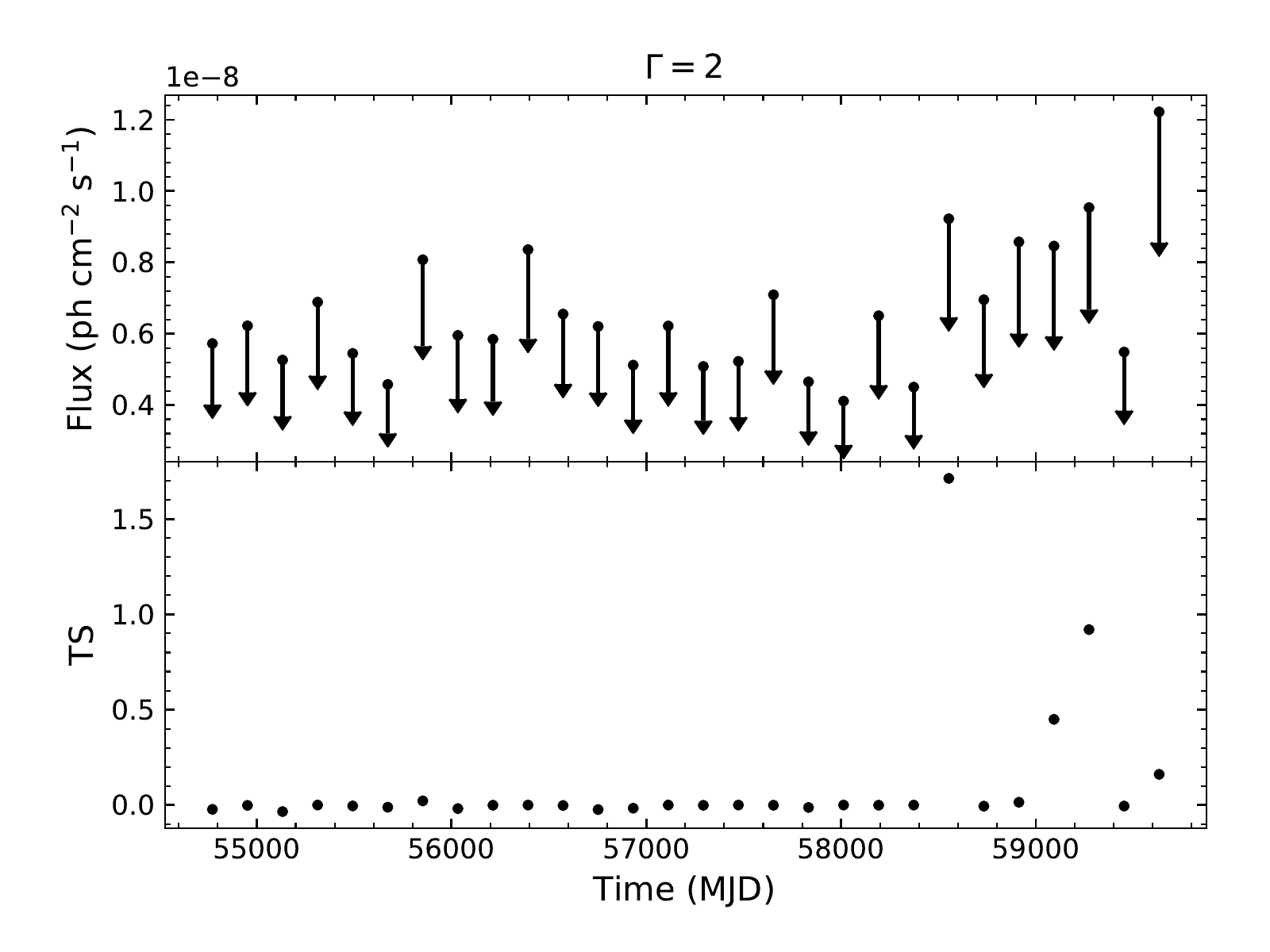}
\includegraphics[scale=0.5]{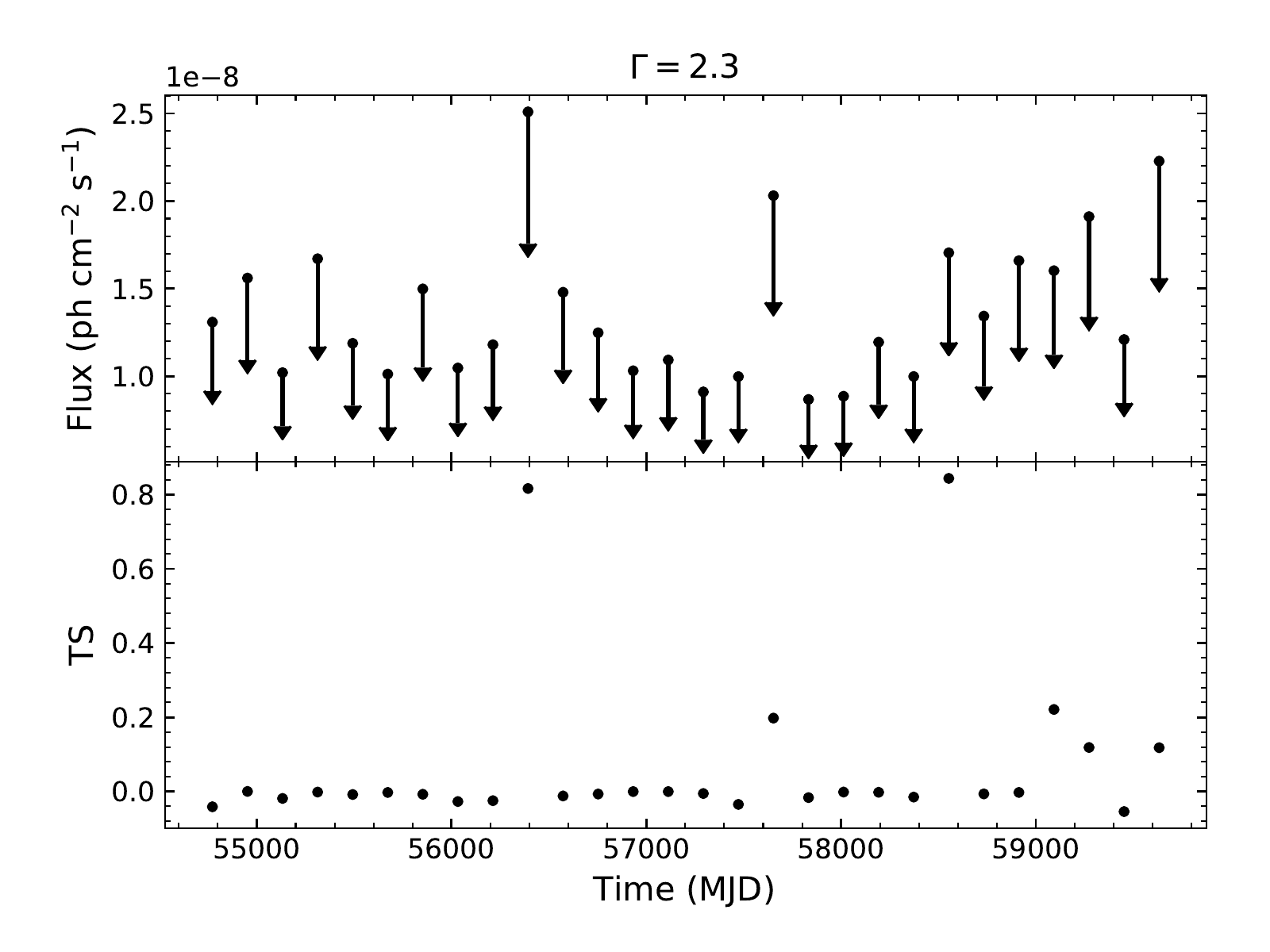} 
\includegraphics[scale=0.5]{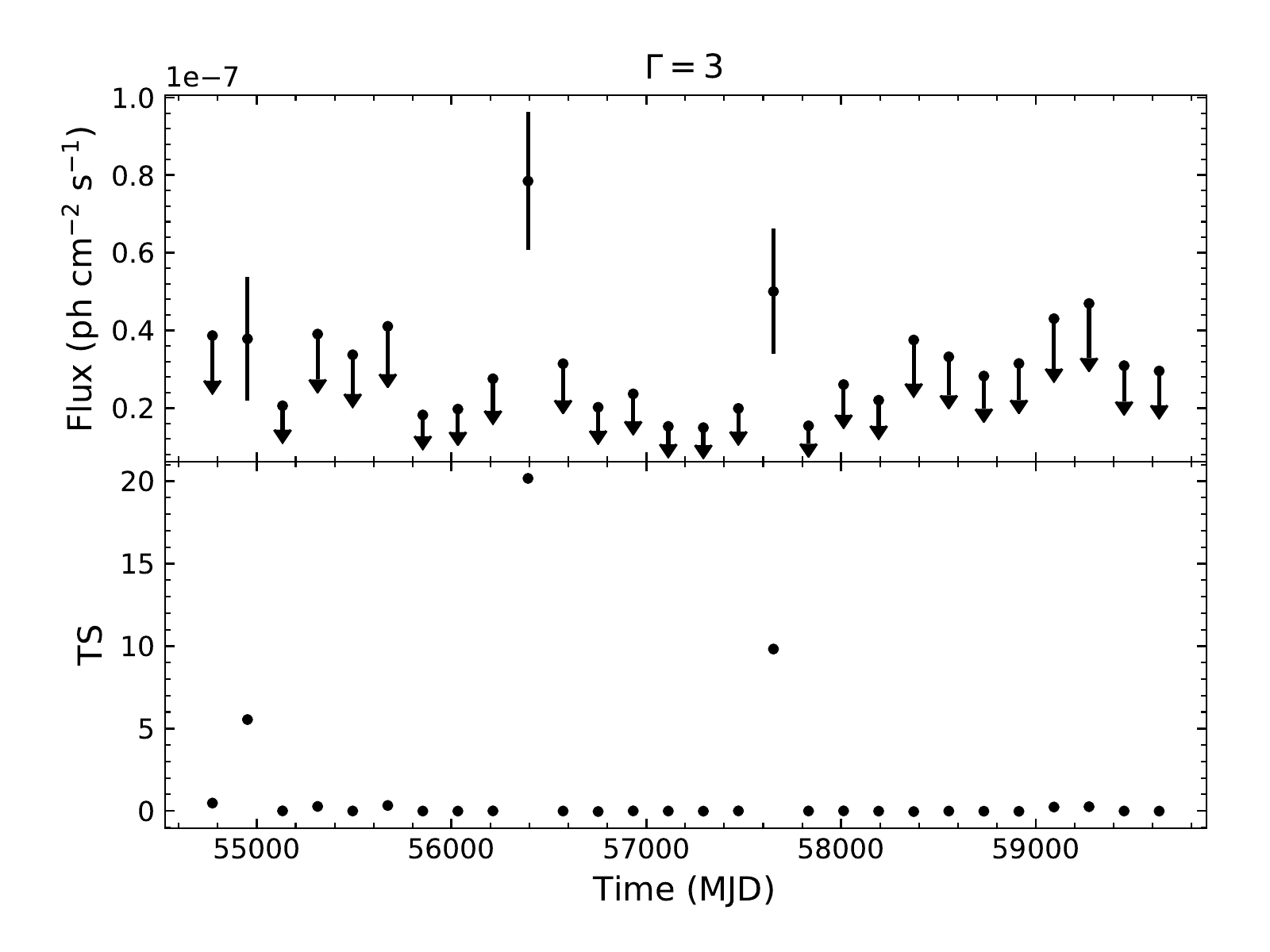} 
\caption{Upper panel: Long-term X-ray light curves of 1A 0535+262 in different energy bands observed by \textit{Swift}/BAT and \textit{MAXI}/GSC as obtained from \protect\url{http://sakamotoagu.mydns.jp/bat_gsc_trans_mon/web_lc/1_Day.php?name=1A_0535+262}. Low three panels: Long-term gamma-ray light curves of 1A~0535+262 with a time binning of 180 days. Spectral index was fixed to 2, 2.3 and 3, respectively. Upper limits at 95\% confidence level are computed for bins with TS$<$4.}
\label{fig:lc}
\end{figure*}

We performed two different types of variability analysis for 1A~0535+262. First, to investigate the long-term gamma-ray flux variability, we computed light curves with a 180-day binning as in \cite{Harvey2022} in the energy range of 0.1$-$300 GeV for all spectral indices used in the spectral analysis (Figure \ref{fig:lc}). 
The full data best-fit source model was used as an input for each time bin and the normalizations of sources within $3^\circ$ around 1A~0535+262 were set free to vary.
Upper limits at 95\% confidence level were calculated when 1A~0535+262 had TS $<4$ in a given time bin. 
We did not see any significant detection in all the time bins when fixing the spectral index to 2 or 2.3. For the light curve with index fixed to 3, there are two bins with TS being about 10 and 20, corresponding to approximately 3$\sigma$ and 4$\sigma$. 
However, these two bins correspond to the period where its X-ray emission was faint according to the X-ray monitoring of 1A~0535+262 (Figure \ref{fig:lc}). Thus, we conclude that no correlation between gamma-ray and X-ray was observed, contrary to what \cite{Harvey2022} has claimed.
Furthermore, the variability significance was computed following the same method used in \cite{3FGL}. 
Only the light curve with index fixed to 3 has a non-negligible significance ($1.7\sigma$ for 27 degrees of freedom), but this is far from declaring a significant variability, which requires usually a significance of at least $3\sigma$. 
Any gamma-ray emission is thus consistent with being steady on a timescale of a few months based on our analysis.

\begin{figure*}
\centering
\includegraphics[scale=0.5]{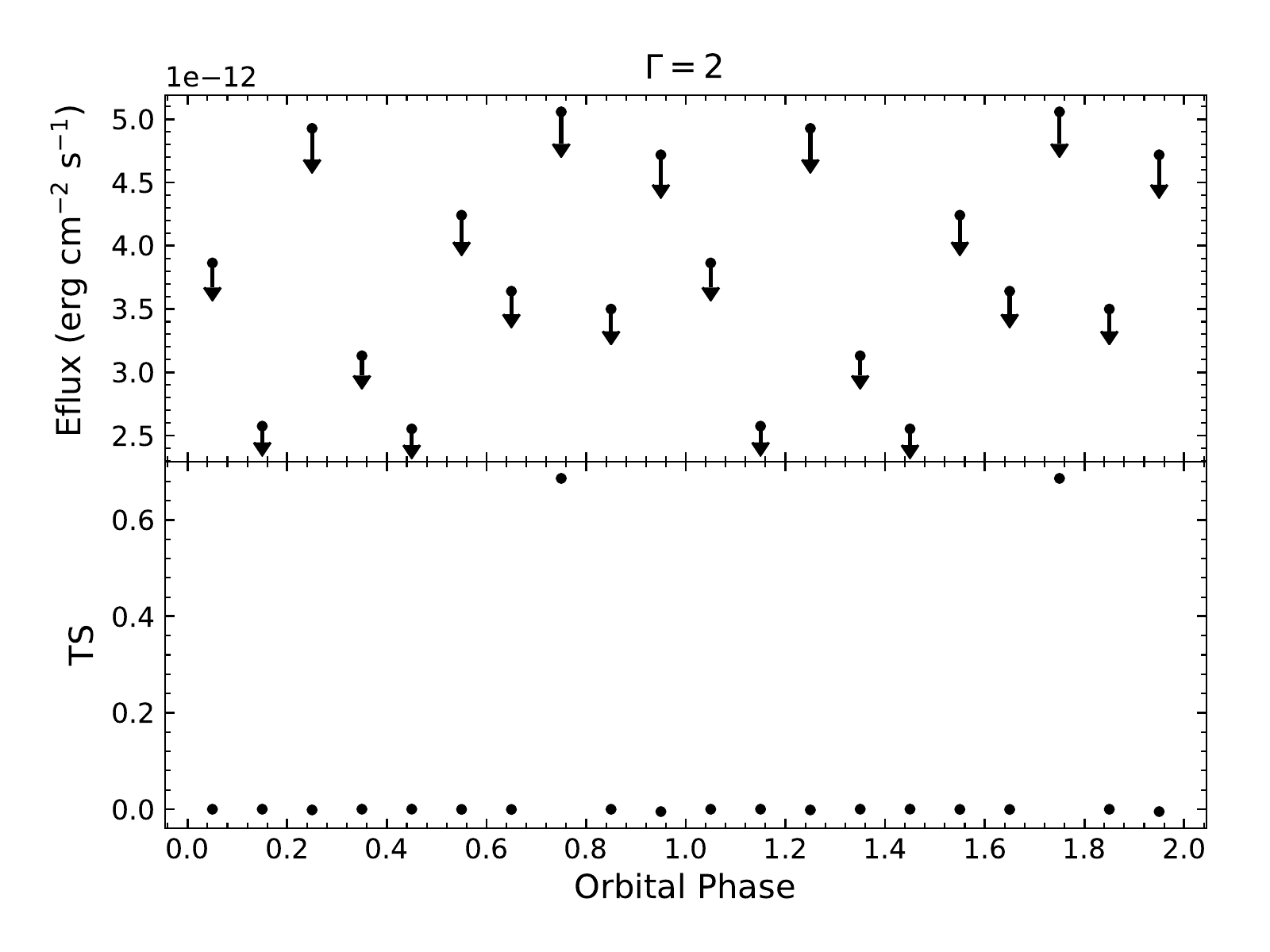}
\includegraphics[scale=0.5]{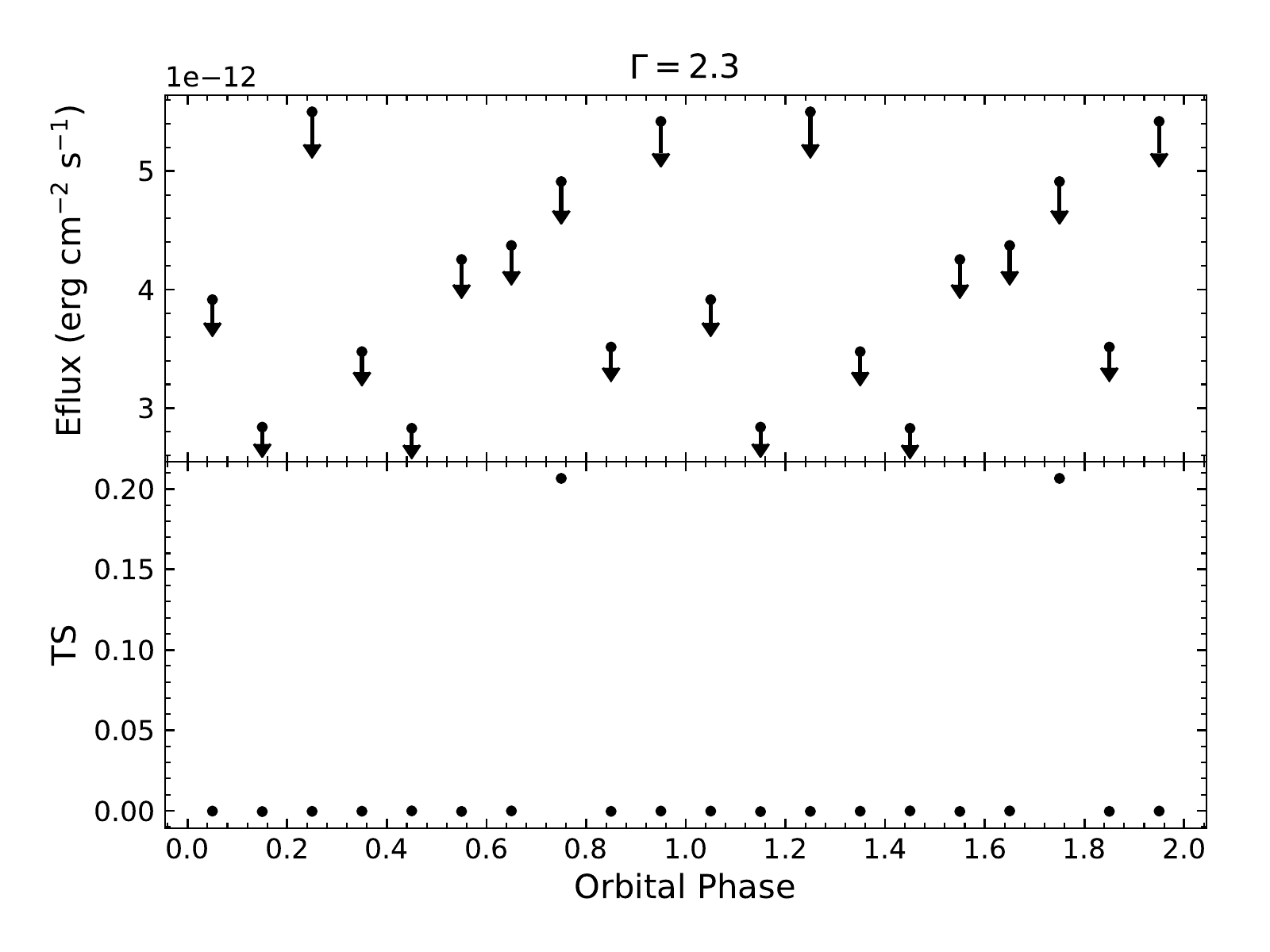} 
\includegraphics[scale=0.5]{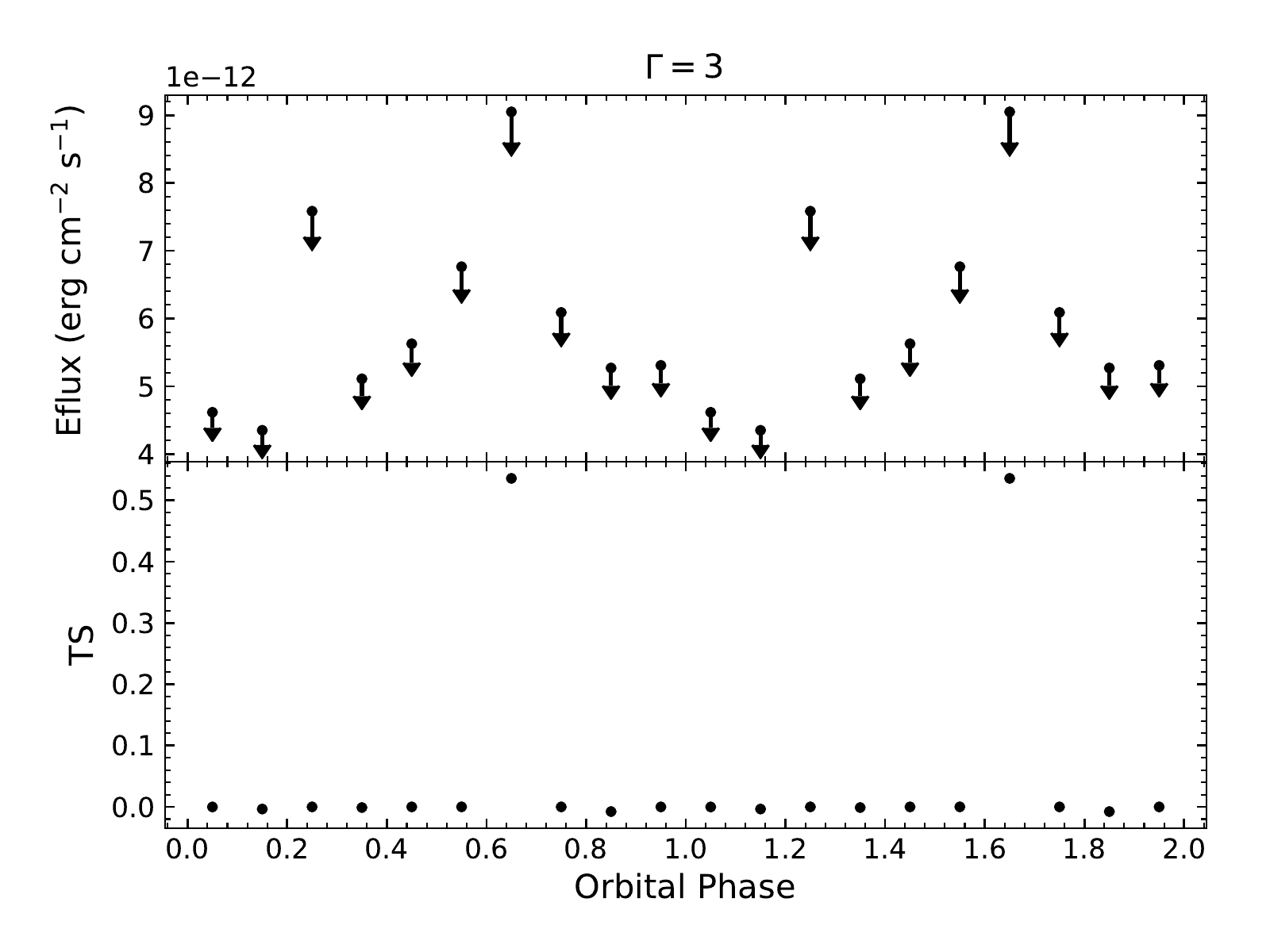} 
\caption{Orbital energy flux variation of 1A~0535+262 with 10 bins per orbit. Spectral index was fixed to 2, 2.3 and 3, respectively. Upper limits at 95\% confidence level are computed for bins with TS$<$4.}
\label{fig:orbit10bins}
\end{figure*}

Since gamma-ray binaries usually exhibit orbital flux variability, we also computed the orbital flux for 1A~0535+262 with 10 bins per orbit and calculated upper limits at 95\% confidence level when 1A~0535+262 had TS $<4$ in a given orbital bin, as was done in \cite{Harvey2022}.
Similar to the long-term light curve, the full data best-fit source model was used as an input for each orbital bin and the spectral index was fixed to 2, 2.3 and 3, respectively. No significant orbital variability was observed.
The orbital light curves are shown in Figure \ref{fig:orbit10bins}.

\subsection{Gamma-ray pulsation search}
\label{pulsation}

We performed a pulsation search using reconstructed LAT photons within $1^\circ $ of 1A~0535+262 in the energy range of 0.1$-$300 GeV. 
The signal significance was qualified using the weighted H-test developed by \cite{Kerr2011}, which is based on the original one proposed by \cite{Jager1989}:
\begin{equation}
    H_{mw}={\rm max}\left[Z^2_{iw}-c\times (i-1)\right], \,\,\,\,\,\,   1\leq i \leq m ,
\end{equation}
where 
\begin{equation}
Z^2_{mw} = \frac{2}{\sum_{i=1}^{N} w_i^2} \times \sum_{k=1}^{m}(\alpha_{wk}^2+\beta_{wk}^2),
\end{equation}
and
\begin{equation}
\alpha_{wk}=\sum_{i=1}^{N} w_i\cos{(2\pi k\phi_{i})}, \;\;\;\;\;\;\beta_{wk}=\sum_{i=1}^{N} w_i\sin{(2\pi k\phi_{i})}.
\end{equation}
Here, $N$ is the total photon number, $\phi_i$ is the pulsar rotational phase and $w_i$ is the photon weight, $m$ is the maximum search harmonic and $c$ is the offset for each successive harmonic. We used the standard value $c=4$ and varied $m$ in our analysis. We verified that taking the standard value $m=20$ did not change the result.

We employed the Simple Weights method descried in \cite{Bruel2019} and \cite{Smith2019} to compute the weight for each photon.
This is considered as a proxy for the probability that the photon comes from 1A~0535+262. 
Assuming that the target source is faint compared to the diffuse background 
and that the background emission is isotropic, for a photon with energy $E$ (in MeV) and angular distance $\Delta \theta$ to the target source, the weight is:
\begin{equation}
    w(E,\Delta \theta) = f(E)\times g(E, \Delta \theta),
\end{equation}
where 
\begin{equation}
    f(E)= \rm exp(-2log_{10}^2(\it E/E_{\rm ref}))
\end{equation}
is the weight at the pulsar position, which depends on the pulsar and background spectra and on the LAT's energy-dependent Point Spread Function (PSF). 
The geometrical factor $g(E, \Delta \theta)$ describes the angular distribution of the gamma-ray photons emitting from a point source and can be written as
\begin{equation}
    g(E, \Delta \theta)= \left(1+\frac{9\Delta \theta^2}{4\sigma_{\rm psf}^2(E)}\right)^{-2},
\end{equation}
where
\begin{equation}
    \sigma_{\rm psf}(E) = \sqrt{p_{0}^2(E/100)^{-2p_1}+p_2^2}
\end{equation}
is the PSF 68\% containment angle with $p_0=5.445, p_1=0.848$ and $ p_2=0.084$ for LAT P305 Pass 8 data \citep{pass8Atwood}.

Defining the reference energy $E_{\rm ref}=10^{\mu_w}$, the H-test versus $\mu_w$ follows a Gaussian distribution \citep{Bruel2019}.
Thus, searching for pulsations means finding the maximum H-test by scanning over $\mu_w$. 
After searching a thousand radio pulsars for possible gamma-ray emission, \cite{Smith2019} indicate that in most cases $\mu_w=3.6$ is a good choice to yield a significant signal. 
We adopted this value in our pulsation search for 1A~0535+262. Considering that this value was found for non-accreting (and many isolated) radio pulsars that have a power law with an exponential cutoff (PLEC) spectrum \citep{Abdo2013}, and thus may not be appropriate for an accreting system that could have a different spectral shape, we also verified that scanning over $\mu_w$ to find the best value does not affect the result significantly. 
We used the ephemerides for different observations (Table \ref{tab:spin}) to phase-fold the gamma-ray photons of 1A~0535+262, restricting the pulsation search range to the validity of the corresponding ephemerides. 
However, no significant pulsation was detected during the individual outbursts. 
%
%
Although a signal of $\sim1.5\sigma$ was hinted for the 2009 outburst, the significance is far from the LAT detection threshold of 4$\sigma$.


\section{Discussion}
\label{discuss}

In general, we obtained very different results compared to \cite{Harvey2022}. We had no detection of either persistent or transient or pulsed gamma-ray emission from 1A~0535+262. 
Although we have two time bins with TS$\sim$20 and $\sim$10, they did not correspond to the X-ray outburst periods. Therefore, our result does not support their conclusion that the gamma-ray emission is correlated with the X-ray outburst of the source and is mostly concentrated in specific orbital bins. Such difference may mainly come from the fitting procedure and whether or not considering the energy dispersion. We normally fitted the sources within $5^\circ$ around 1A~0535+262 in the spectral fit and orbital flux variability study, and within $3^\circ$ for the long-term light curves, while they fitted only $1^\circ$ around 1A~0535+262. In addition, the spectral index was fixed to different values to account for possible emission mechanisms in our study, while they let the index free to vary. Actually, for the long-term variability, no detailed fitting procedure information was found in \cite{Harvey2022}, making a detailed comparison difficult. 


Unlike those cases when a detection is found from an astrophysical source, the possible reasons for a non-detection are essentially unlimited.
Gamma-ray binary systems such as LS I +61 303 are radio sources, and their X-ray spectra contain a significant non-thermal component. Recently, radio pulsations were detected from this system \citep{Weng2022}. Thus, it is worth noting that LS I +61 303 is probably not a significantly accreting system, and thus it likely has a different acceleration and emission mechanism than that acting in 1A 0535+26, if there is any in the latter.
On the other hand, in 1A~0535+262, the absence of a significant quiescent radio emission that could later be associated with non-thermal processes may simply indicate that leptons are not sufficiently accelerated there.
This observation had earlier promoted hadronic models. 
From our results, though, we can also rule out hadronic production acting according to a mechanism originally proposed by \cite{Cheng1991}, where a proton beam accelerated in a magnetospheric electrostatic gap impacts the transient accretion disk. 
This model was applied to 1A~0535+262 by \cite{Romero2001}, \cite{Anchordoqui2003} and \cite{Orellana2007}. 
Particularly in the latter paper the theoretical flux prediction of $3.8\times 10^{-8} \, \rm ph\ cm^{-2}$ \citep[derived by extrapolating the result of][]{Orellana2007}, 
i.e., a gamma-ray luminosity of about 10$^{33}$ erg s$^{-1}$ at 0.3 TeV at the end of giant outbursts to the \textit{Fermi}-LAT energy range \citep[see][]{Acciari2011} is above our limit which is $\sim$(2.3$-$4.7)$\times 10^{32}\, \rm erg \, s^{-1}$ by one order of magnitude (and should have been detected earlier on in the mission) depending on different spectral indices assumed. Taking the X-ray luminosity (2$-$150 keV) reported for the largest 2020 outburst \citep{Kong2021} to be 1.2$\times$10$^{38}\, \rm erg \, s^{-1}$, the ratio of $L_{\rm \gamma}$ to $L_{\rm X}$ is then calculated to be (1.9$-$3.9)$\times10^{-6}$.

It remains to be seen, of course, whether the flux was overestimated but the mechanism is still viable, or whether a longer integration might lead to low-flux quiescent emission or to an occasional detection. 
None of these possibilities has happened in the long integration time we have analyzed. 
At least with the current gamma-ray sensitivity, 1A~0535+262 is not a gamma-ray source. 
With our current limits, there appears to be no reasonable combination of the hadronic model parameters to still accommodate a persistent gamma-ray flux.

Another interpretation could simply be that none of the possible shocks in the system is energetic enough to produce sufficiently accelerated particles able to emit gamma-rays. Or that if they do, the radiation produced is absorbed due to the matter in the surroundings, see e.g., the discussion in \cite{Orellana2007}.
Regarding the latter, it would be reasonable to expect that any absorption would be quite variable. 
Thus if the emission is produced in the system at all, it would be unlikely that it is absorbed all the time.
In addition, secondary electrons (and positrons) from the pair production process would generate gamma rays at MeV–GeV energies, \citep[see e.g.,][]{Bednarek1997,Bednarek2006,Sierpowska-Bartosik2008}. 
These could be attenuated by X-ray photons, but probably not fully attenuated after the X-ray peak has passed.

Giant (or Type II) outbursts, though they are rare and unrelated to the orbital cadence, may have X-ray luminosities close to the Eddington limit. 
They are likely associated with the formation of a transient accretion disk.
Recently, \cite{Eijnden2020} detected a radio counterpart during the 2020 outburst.
This was the first time that a coupled increase in X-ray and radio flux was seen in 1A~0535+262 and shows that the radio emission may relate to the accretion state.
This would be similar to the behaviour seen in the transient Be X-ray binary Swift J0243.6+6124 \citep{Eijnden2018}.
\cite{Bednarek2009,Bednarek2009b} proposed that it is possible that HMXBs produce gamma-ray emission during accretion periods. 
The possibility that particle acceleration can take place even when mass accretion is going on is supported by some observational results: the hinted gamma-ray emission from SAX J1808.4$-$3658 \citep{Emma2016}, the gamma-ray emission found from the sub-luminous state of the transitional pulsars (as quoted in the introduction) and interpreted as propeller emission
\cite[see e.g.,][]{Papitto2014,Papitto2015} or a mini pulsar wind nebula (\cite{Papitto2019}, see also \cite{Veledina2019})
and finally, 
the discovery of optical and ultraviolet pulsed emission from the accreting millisecond pulsar SAX J1808.4$-$3658 \citep{Ambrosino2021}.
However, the neutron star Eddington luminosity  ($L_{\rm Edd} \sim 1.8 \times 10^{38}$ erg s$^{-1}$) is many orders of magnitude above our upper limits, pointing to a very inefficient mechanism, if at play at all.
Similarly to the case of other transients, for instance, the Be X-Ray Binary 4U 1036$-$56 (RX J1037.5$-$5647), which could be associated to \textit{AGILE} transients,
we cannot discard that a low level of gamma-ray flux is emitted at lower energies, below 100 MeV \citep[see][and in particular their figure 6 for an associated discussion]{Li2012}. 
This remains to be tested with the advent of MeV missions such as AMEGO \citep{McEnery2019}, e-ASTROGAM \citep{Angelis2018}, or COSI \citep{Beechert2022}.

\section{Conclusions}
\label{conclude}

We have searched for gamma-ray emission and pulsations from 1A~0535+262 using more than 13 years of \textit{Fermi}-LAT data. 
Neither persistent nor transient nor pulsed gamma-ray emission has been detected significantly in our study for the whole data set or during the X-ray outbursts that 1A~0535+262 has experienced since the launch of the \textit{Fermi} satellite. 
Upper limits on the luminosity at the 95\% confidence level were derived to be around (2.3$-$4.7)$\times 10^{32}\, \rm erg \, s^{-1}$ depending on different spectral indices assumed, the deepest ones to date. The emission of 1A~0535+262 is considered to be consistent with being steady on a timescale of a few months. 
Although two time bins in the long-time light curve hint to have gamma-ray emission at roughly 3 and 4$\sigma$, they occurred when the source was faint in X-rays. 
Thus, no correlation between gamma-ray and X-ray activities was observed based on our result.
In addition, we did not see any significant orbital gamma-ray variation. 
We conclude that 1A~0535+262 is not a gamma-ray emitter at the level of the current gamma-ray sensitivity.

\acknowledgments
The \textit{Fermi} LAT Collaboration acknowledges generous ongoing support from a number of agencies and institutes that have supported both the development and the operation of the LAT, as well as scientific data analysis. These include the National Aeronautics and Space Administration and the Department of Energy in the United States; the Commissariat \`a l'Energie Atomiqueand and the Centre National de la Recherche Scientifique/Institut National de Physique Nucl\'eaire et de Physique des Particules in France; the Agenzia Spaziale Italiana and the Istituto Nazionale di Fisica Nucleare in Italy; the Ministry of Education, Culture, Sports, Science and Technology (MEXT), High Energy Accelerator Research Organization (KEK), and Japan Aerospace Exploration Agency (JAXA) in Japan; and the K.~A.~Wallenberg Foundation, the Swedish Research Council, and the Swedish National Space Board in Sweden. Additional support for science analysis during the operations phase is gratefully acknowledged from the Istituto Nazionale di Astrofisica in Italy and the Centre National d'\'Etudes Spatiales in France. 

This work was performed in part under DOE Contract DE-AC02-76SF00515. The authors are supported by the National Natural Science Foundation of China through grants U1938103, 12041303, 12173103, U2038101, 11733009. WZ and DFT work have been supported by the grant PID2021-124581OB-I00 funded by MCIN/AEI/10.13039/501100011033
and by the Spanish program Unidad de Excelencia María de
Maeztu CEX2020-001058-M. DFT acknowledges as well USTC and the Chinese Academy of Sciences International Presidential Fellowship Initiative 2021VMA0001.  
This research has made use of the SIMBAD database, operated at CDS, Strasbourg, France.

\bibliography{A0535}{} 
\bibliographystyle{aasjournal}

\begin{table}
 \scriptsize
\begin{center}
\begin{threeparttable}
\caption{\textit{Fermi}-LAT spectral analysis results} 
\label{tab:fermiflux}
\begin{tabular}{lcccc}
 \toprule
 Period$^{a}$  & Time Range &  Spectral Index  &  TS   &  Energy Flux Upper Limit$^{b}$  \\[0.8ex]
         & (MJD)      &                  &       &  $(10^{-12}\, \rm erg \, cm^{-2}\,s^{-1})$  \\[0.8ex]
\hline
  \multicolumn{5}{c}{Whole dataset}     \\  
   \hline   
  full & 54682-59739 &  $2.0$ & $0.0$ & $0.6$  \\ 
  full & 54682-59739 &  $2.3$ & $0.0$ & $0.7$  \\ 
  full & 54682-59739 &  $3.0$ & $0.0$ & $1.2$  \\ 
   \hline   
   \multicolumn{5}{c}{Stacked outbursts}     \\  
   \hline   
  Rising+Falling & 55040-59207 &  $2.0$ & $0.0$ & $11.9$  \\ 
  Rising+Falling & 55040-59207 &  $2.3$ & $0.0$ & $9.3$  \\ 
  Rising+Falling& 55040-59207 &  $3.0$ & $0.0$ & $20.9$  \\ 
   \hline   
   \multicolumn{5}{c}{The 2009 double-peaked outburst}     \\  
   \hline   
  Rising+Falling & 55040-55070 &  $2.0$ & $0.4$ & $34.9$  \\ 
  Rising+Falling & 55040-55070 &  $2.3$ & $0.4$ & $27.3$  \\ 
  Rising+Falling & 55040-55070 &  $3.0$ & $0.0$ & $21.9$  \\ 
   \hline   
   \multicolumn{5}{c}{The 2009 giant outburst}     \\  
   \hline   
  ALL & 55165.9-55249.6 &  $2.0$ & $0.0$ & $9.9$  \\ 
  ALL & 55165.9-55249.6 &  $2.3$ & $0.0$ & $10.5$  \\ 
  ALL & 55165.9-55249.6 &  $3.0$ & $0.0$ & $16.5$  \\ 
  Rising+Falling & 55165.9-55193.6 &  $2.0$ & $0.0$ & $27.9$  \\ 
  Rising+Falling & 55165.9-55193.6 &  $2.3$ & $0.0$ & $24.7$  \\ 
  Rising+Falling & 55165.9-55193.6 &  $3.0$ & $0.0$ & $28.4$  \\ 
  Rising & 55165.9-55177.6 &  $2.0$ & $0.0$ & $49.3$  \\ 
  Rising & 55165.9-55177.6 &  $2.3$ & $0.0$ & $45.8$  \\ 
  Rising & 55165.9-55177.6 &  $3.0$ & $0.0$ & $45.4$  \\ 
  Falling & 55178.4-55193.6 &  $2.0$ & $0.0$ & $39.4$  \\ 
  Falling & 55178.4-55193.6 &  $2.3$ & $0.0$ & $31.7$  \\ 
  Falling & 55178.4-55193.6 &  $3.0$ & $0.0$ & $29.5$  \\ 
  Apastron & 55199.4-55216.6 &  $2.0$ & $0.0$ & $23.6$  \\ 
  Apastron & 55199.4-55216.6 &  $2.3$ & $0.0$ & $18.1$  \\ 
  Apastron & 55199.4-55216.6 &  $3.0$ & $0.0$ & $18.6$  \\ 
  Periastron & 55230.4-55249.6 &  $2.0$ & $0.0$ & $44.9$  \\ 
  Periastron & 55230.4-55249.6 &  $2.3$ & $0.0$ & $39.3$  \\ 
  Periastron & 55230.4-55249.6 &  $3.0$ & $0.0$ & $34.5$  \\ 
   \hline   
   \multicolumn{5}{c}{The 2011 giant outburst}     \\  
   \hline   
  Rising+Falling & 55600-55645 &  $2.0$ & $0.0$ & $14.6$  \\ 
  Rising+Falling & 55600-55645 &  $2.3$ & $0.0$ & $13.4$  \\ 
  Rising+Falling & 55600-55645 &  $3.0$ & $0.0$ & $19.6$  \\ 
  Rising & 55600-55617 &  $2.0$ & $0.0$ & $33.0$  \\ 
  Rising & 55600-55617 &  $2.3$ & $0.0$ & $30.7$  \\ 
  Rising & 55600-55617 &  $3.0$ & $0.0$ & $40.0$  \\ 
  Falling & 55618-55645 &  $2.0$ & $0.0$ & $30.9$  \\ 
  Falling & 55618-55645 &  $2.3$ & $0.0$ & $30.6$  \\ 
  Falling & 55618-55645 &  $3.0$ & $0.0$ & $20.0$  \\ 
   \hline   
   \multicolumn{5}{c}{The 2020 giant outburst}     \\  
   \hline   
  Rising+Falling & 59159-59207 &  $2.0$ & $0.0$ & $33.4$  \\ 
  Rising+Falling & 59159-59207 &  $2.3$ & $0.0$ & $27.2$  \\ 
  Rising+Falling & 59159-59207 &  $3.0$ & $0.0$ & $31.4$  \\ 
  Rising & 59159-59172.5 &  $2.0$ & $4.2$ & $131.5$  \\ 
  Rising & 59159-59172.5 &  $2.3$ & $3.2$ & $91.8$  \\ 
  Rising & 59159-59172.5 &  $3.0$ & $0.5$ & $90.0$  \\ 
  Falling & 59173-59207 &  $2.0$ & $0.0$ & $24.2$  \\ 
  Falling & 59173-59207 &  $2.3$ & $0.0$ & $22.0$  \\ 
  Falling & 59173-59207 &  $3.0$ & $0.0$ & $30.5$  \\ 


\bottomrule
\end{tabular}
\begin{tablenotes}
     \scriptsize
      \item{}{$^{a}$ Full: the whole dataset used in this work; ALL: the dataset including the rising, falling, apastron and periastron portions of the outburst.} 
      \item{}{$^{b}$ The upper limits are given at the 95\% confidence level in the energy range of 0.1$-$300 GeV. }
      \vspace{0.5cm} 
\end{tablenotes}
\end{threeparttable}
\end{center}
\end{table}

\begin{table*}[h!]
	\begin{center}
		\caption{Parameters of the spin evolution of 1A 0535+026. $\nu_i$ is the $i$th order derivative of the frequency. 
		}
		\label{tab:spin}
		\begin{tabular}{c|c|c|c|c} 
			\hline
			Parameters                          & 2009 double outburst$^{a}$    & 2009 outburst$^{a}$        & 2011 outburst$^{b}$         & 2020 outburst$^{c}$ \\
		\hline
		Epoch (MJD)                             & 55040   & 55166.99     & 55616.202    & 59170         \\
		$T_{\rm start}$ (MJD)                   & 55040   & 55166.99    & 55608        & 59159.15      \\
		$T_{\rm stop}$ (MJD)                    & 55070   & 55201       & 55637        & 59207.92       \\
		$\nu_0$($\rm 10^{-3}\,Hz$)              & 9.66041(4)  & 9.6618(1)   & 9.6793(1)    & 9.66045(2)      \\
		$\nu_1$($\rm 10^{-12}\,Hz\,s^{-1}$)    &   0.67(3)    & -4.6(6)  & 6.43(5)    & 19.27(4)    \\
		$\nu_2$($\rm 10^{-17}\,Hz\,s^{-2}$)    &      & 3.3(0.3)     & 0.121(7)   & 1.17(4)    \\
		$\nu_3$($\rm 10^{-23}\,Hz\,s^{-3}$)    &      &    -4.8(6)      & -1.43(9)   & -5.8(2)    \\
		$\nu_4$($\rm 10^{-29}\,Hz\,s^{-4}$)    &      &     3(1)          & 1.67(5)   & -2.8(9)   \\
		$\nu_5$($\rm 10^{-36}\,Hz\,s^{-5}$)    &      &    -8(6)          &         & 737(7)     \\
		$\nu_6$($\rm 10^{-39}\,Hz\,s^{-6}$)    &      &                  &         & -1.6(2)      \\
		$\nu_7$($\rm 10^{-45}\,Hz\,s^{-6}$)    &      &                  &         & -5(2)      \\
		$\nu_8$($\rm 10^{-50}\,Hz\,s^{-6}$)    &      &                  &         & 3(1)      \\
		$\nu_9$($\rm 10^{-56}\,Hz\,s^{-6}$)    &      &                  &         & -8(2)      \\
		$\nu_{10}$($\rm 10^{-62}\,Hz\,s^{-6}$)    &      &                 &        & 7(1)      \\

		\hline
		\end{tabular}
	\begin{tablenotes}
     \scriptsize
     \item $^{a}$: Derived form the {\it Fermi}/GBM monitoring (see text). 
     \item $^{b}$: Adopted from \cite{Sartore2015}.
     \item $^{c}$: Derived from {\it Insight}-HXMT observations \citep[][see text]{Wang2022}.
    \end{tablenotes}
	\end{center}
\end{table*}

\end{document}